\documentclass[letterpaper,showpacs,floatfix,aps]{revtex4}

\usepackage{amssymb}

\def \beq {\begin{equation}}
\def \eeq {\end{equation}}

\begin{document}

\title{Homogeneous cosmologies and the Maupertuis-Jacobi principle}

\author{Luciana A. Elias}
\author{Alberto Saa}
 \email{asaa@ime.unicamp.br}
  \affiliation{
  Departamento de Matem\'atica Aplicada, \\
  IMECC -- UNICAMP, C.P. 6065, \\ 
  13083-859 Campinas, SP, Brazil.}


\begin{abstract}
A recent work showing that homogeneous and isotropic cosmologies involving
scalar fields are equivalent to the geodesics of certain effective manifolds is 
generalized to the  non-minimally coupled and anisotropic cases. 
As the Maupertuis-Jacobi principle in classical mechanics, such result
permits us to infer some dynamical properties of   cosmological models
from the geometry of the associated effective manifolds, allowing us to
go a step further in the study of cosmological dynamics.
By means of some explicit examples, we show how the geometrical analysis
can simplify considerably the dynamical analysis of cosmological
models.
\end{abstract}
\maketitle

\section{Introduction}

The dynamical system approach has a prominent  role in modern cosmology. 
Any candidate to a realistic cosmological model must exhibit certain qualitative dynamical behaviors
as, among others,   robustness against small perturbations (to avoid fine tunings
problems, see \cite{fine} for a recent review),   
specific kinds of phase space attractors (to describe, for instance,
the recent accelerated expansion phase  of the universe\cite{accel,Peebles}), and some
classes of solutions with determined 
intermediate time behavior (for instance, tracker solutions\cite{Peebles,tracker}, candidates to 
explain the Cosmic Coincidence or Why Now? problem). In a 
  recent work\cite{townsend},   Townsend and Wohlfarth present  an interesting and promising tool
for the dynamical analysis of cosmological models. Essentially, they show that
the equations of motion of homogeneous and isotropic cosmological models with
multiple minimally coupled scalar fields, self interacting through an arbitrary potential, 
do indeed correspond to the geodesic equations of a certain effective Lorentzian 
manifold. Such correspondence, closely related to the Maupertuis-Jacobi principle
of classical mechanics\cite{Kozlov}, permits us to infer some of the dynamical properties of
a given cosmological model  from the underlying geometry of the associated effective manifold,
allowing one to go a step further in the dynamical analysis of realistic  cosmological
models.  Townsend and Wohlfarth, indeed, apply their own formalism to identify 
asymptotic accelerated expansion phases in models with exponential potentials\cite{townsend}.
As we will see, such geometrical considerations
can simplify considerably the dynamical study of  cosmological
models.

Here, we show that the work of Townsend and Wohlfarth can be
largely extended, allowing the inclusion of non-minimally coupled scalar fields and also the
anisotropic case, opening many new  possibilities for the dynamical characterization
of such cosmological models. Our main results and some applications are
presented in the next sections, after the following
brief introduction to the 
Maupertuis-Jacobi principle. 

\subsection{The Maupertuis-Jacobi principle}

The Maupertuis-Jacobi principle in classical mechanics\cite{Kozlov} establishes that the dynamics
of a given system can be viewed as geodesic motions in an associated Riemannian manifold. 
In order to recall it briefly,  
let us consider a classical mechanical
system with $N$ degrees of freedom described by the Lagrangian
\beq
\label{lagr}
{  L}(q,\dot{q}) = \frac{1}{2} g_{ij}(q)\dot{q}^i \dot{q}^j -
V(q),
\eeq
where $i,j=1, 2,\dots, N$,
the dot stands
for differentiation with respect to the time $t$, and
$g_{ij}$ is the Riemannian
metric on the  $N$-dimensional configuration space $\mathcal M$. All the
quantities here are assumed to be smooth. The Euler-Lagrange
equations of (\ref{lagr}) can be written as
\beq
\label{EL}
\ddot{q}^i + \Gamma^i_{jk} \dot{q}^j \dot{q}^k = -
g^{ij}\partial_j V(q),
\eeq
where $\Gamma^i_{jk}$ is the Levi-Civita connection for the
metric $g_{ij}$.
The Hamiltonian of the system described by (\ref{lagr})
\beq
\label{ham}
{\cal H}(q,p) = \frac{1}{2} g^{ij}(q){p}_i {p}_j +
V(q),
\eeq
with $p_i=g_{ij}\dot{q}^j$,
is obviously a constant of motion, namely the total energy.
For a fixed energy $E$, the trajectories in the $2N$-dimensional phase-space $(q^i;p_j)$
are confined to the hypersurface
$E = \frac{1}{2} g^{ij}{p}_i {p}_j + V(q).$
The admissible region for the trajectories in the configuration
space is given by
\beq
{\mathcal D}_E = \{q\in{\mathcal M} : V(q) \le E \},
\eeq
which can be either bounded or unbounded, connected
or not. 
If the potential $V$ has no critical points on the boundary
$\partial {\mathcal D}_E $, then
$\partial {\mathcal D}_E $ is a $N-1$ dimensional submanifold
of $\mathcal M$.
We can easily see that if a trajectory reaches the boundary
  at a point $q_0$, its velocity at
this point vanishes and the trajectory approach
or depart from $q_0$ perpendicularly
to  $\partial {\mathcal D}_E $. In particular,
there is no allowed trajectory along the boundary.

One can show that the equations of
motion (\ref{EL}) are, in the interior of
${\mathcal D}_E$,  fully equivalent to the geodesic equation
of the ``effective''
Riemannian geometry on $\mathcal M$ defined from the Jacobi
metric\cite{Kozlov} 
\beq
\label{jacobi}
\hat{g}_{ij}(q) = 2(E-V(q)) g_{ij}(q).
\eeq
The   geodesic
equation in question is given by
\beq
\label{geo}
\hat{\nabla}_u u =
\frac{d^2 q^i}{ds^2} + \hat{\Gamma}^i_{jk}
\frac{dq^j}{ds} \frac{dq^k}{ds} = 0,
\eeq
where
$u=dq^i/ds$ is the tangent vector along the geodesic,
$\hat{\nabla}$ and
$ \hat{\Gamma}^i_{jk}$ are, respectively, the covariant derivative
and the Levi-Civita connection
for the Jacobi metric $\hat{g}_{ij}$, and $s$ is a parameter
along the geodesic obeying
\beq
\label{para}
\frac{ds}{dt} = 2 (E-V(q)).
\eeq
As for any classical topic, there is a vast literature on the  Maupertuis-Jacobi principle.
We notice only that, motivated
by the celebrated result due to
Anosov stating that the geodesic flow in a compact manifold with all
sectional curvatures negative at every point is chaotic\cite{Kozlov}, 
the Maupertuis-Jacobi principle
  has been 
recently invoked for the study of chaotic dynamics. (See, for instance,
\cite{chaos} and the references therein).

Townsend and Wohlfarth consider homogeneous and
isotropic cosmological models with $N$ self-interacting minimally coupled 
scalar fields $\phi^\alpha$ taking their values in
a Riemannian target space endowed with a metric $G_{\alpha\beta}$. The corresponding actions is
\beq
\label{act}
S=\int d^Dx \sqrt{-g}\left(R - g^{ij}G_{\alpha\beta}(\phi)\partial_i\phi^\alpha \partial_j\phi^\beta - 2 V(\phi) \right).
\eeq 
where  $R$ stands for
the scalar curvature of the $D$-dimensional spacetime metric  $g_{ij}$. 
By considering the Friedman-Robertson-Walker homogeneous and isotropic metric
\beq
\label{RW}
ds^2  = -dt^2 + a^2(t)d\Sigma_\kappa^2,
\eeq
where $\Sigma_\kappa$ represents the $(D-1)$-dimensional spatial sections of constant curvature $\kappa$,
they showed  that the equations
of motion associated to the action (\ref{act}) do indeed correspond to the geodesics of a certain
 effective Jacobi  (pseudo) metric on a   Lorentzian manifold. For the spatially flat
case ($\kappa=0$), for instance, 
the geodesics corresponding to the 
equations of motion derived from (\ref{act}) are   timelike, null,
or spacelike   according, respectively, if $V>0$, $V=0$, or $V<0$. Such results have
been already 
applied to the dynamical study of the models governed by actions of the type (\ref{act}), 
see \cite{appl}. For $\kappa\ne 0$, Townsend and Wohlfarth had to
introduce some higher dimensional effective
manifolds to accomplish their analysis. 
As we will see,  such higher dimensional manifolds are unnecessary,  all values of $\kappa$
can be eventually treated in the same framework.

Applications of the
Maupertuis-Jacobi principle to the field equations obtained from   Hilbert-Einstein like actions
have also a long history. Non-homogeneous and anisotropic cases were considered in \cite{nh}.
Applications involving distinct differential
spaces instead of differential manifolds were discussed in
\cite{SHS}. Non-minimally coupled   fields, however, have not been considered so far.

\section{Non-minimally coupled homogeneous and isotropic cosmologies}

Non-minimally coupled scalar fields are quite common in cosmology. In particular, they have been invoked
recently to describe dark energy (see \cite{gunzig} and \cite{FS}, and the references therein, for,
respectively,   models using conformal coupling and   more general ones. See 
also \cite{fine,Peebles}.).
The non-minimal coupling generalization of (\ref{act}) we consider here
is
\beq
\label{act1}
S=\int d^4x \sqrt{-g}\left(F(\phi)R - g^{ij}G_{\alpha\beta}(\phi)\partial_i\phi^\alpha \partial_j\phi^\beta - 2 V(\phi) \right).
\eeq
Integrating by parts the action (\ref{act1}), taking into account that 
 $R=6\dot{H}+12H^2  + 6\kappa/a^2$ for the metric (\ref{RW}), with $H=\dot{a}/a$, we
obtain the following Lagrangian 
\beq
\label{l1}
L(a,\dot{a},\phi_\alpha,\dot{\phi}^\alpha) = a^3\left(- 6 H^2 F(\phi) - 6H  \dot{\phi}^\alpha\partial_\alpha F(\phi)  
   + G_{\alpha\beta}(\phi)\dot{\phi}^\alpha \dot{\phi}^\beta + 6\kappa \frac{F(\phi)}{a^2}- 2V(\phi) \right)  
\eeq
on the $(N+1)$-dimensional configuration space spanned by $(a,\phi^\alpha)$. 
Although we will discuss here only the $D=4$ case, the $D$-dimensional one follows straightforwardly
by using that $R=2(D-1)\dot{H}+D(D-1)H^2  + (D-1)(D-2)\kappa/a^2$  in order to derive (\ref{l1}).
By introducing the following Lorentzian metric
\beq
\label{m}
G_{AB}(a,\phi^\alpha) = \left(
\begin{array}{cc}
 -6aF& {-3a^2\partial_\beta F} \\ 
 {-3a^2\partial_\alpha F}&  a^3G_{\alpha\beta}  
\end{array}  \right)
\eeq
on the configuration space (upper case roman indices run over $1\dots N+1$), 
the Lagrangian (\ref{l1}) can be 
cast in the form
\beq
\label{l2}
L(\phi^A,\dot{\phi}^A) = G_{AB}(\phi)\dot{\phi}^A \dot{\phi}^B - 2V_{\rm eff}(\phi^A),
\eeq
where $\phi^A= (a,\phi^\alpha)$ and 
\beq
V_{\rm eff}(\phi^A) = a^3V(\phi) - 3\kappa a{F(\phi)}. 
\eeq
It is evident the similarity between   (\ref{lagr})  and (\ref{l2}), 
provided that $\det G_{AB} \ne 0$, which we discuss below.
Before, let us recall that our manipulations imply that
all solutions of the Euler-Lagrange of (\ref{act1}) are also solutions of the Euler-Lagrange equations
of (\ref{l1}), but not the converse. Einstein equations form a constrained system, the solutions
of (\ref{act1}) correspond, indeed, to a subset of the solutions of (\ref{l1}), as one can realize by
considering the  Hamiltonian associated to (\ref{l1})
\beq
\label{h2}
{\cal  H}(\phi^A, \pi_A) = G^{AB}\pi_A \pi_B + 2V_{\rm eff}(\phi^A),
\eeq
which is a constant of motion,
where $\pi_A = G_{AB}\dot\phi^B$.  The Euler-Lagrange equations of
(\ref{act1}), on the other hand, implies that ${\cal H}=0$ (the so-called energy constraint). Hence, we must
bear in mind that the relevant solutions of our original problem correspond, in fact, 
to the ${\cal H}=0$ subset of the
dynamics governed by (\ref{l1}). 

A proper interpretation of (\ref{m}) as a metric tensor does require the
essential assumption  $\det G_{AB} \ne 0$. In order to grasp it actual 
meaning and implications, let us restrict ourselves to  the $N=1$ and $G=1$ case, 
for which 
\beq
\label{det}
\det G_{AB} = -6a^4\left(F(\phi) + \frac{3}{2}\left( F'(\phi) \right)^2 \right).
\eeq
The vanishing of the quantity between parenthesis is known for a long time
to be associated with the existence of some
severe and 
unavoidable dynamical singularities\cite{sing1} (see also \cite{sing}), which render 
the associated cosmological 
model unphysical. Here, the vanishing of (\ref{det})  also leads generically to an actual geometrical singularity.
Thus,
the assumption of $\det G_{AB} \ne 0$ assures that the model in question is free
of these singularities, geometrical and dynamical, a 
basic requirement for any realistic model.
This can be viewed as the first application of the formalism; 
we will return to this issue in the last section. 
Note that one of the most
interesting  peculiarities  of the conformal coupling $(N=1, F(\phi)=1-\phi^2/6)$ is that it 
can always
evade the $\det G_{AB} = 0$
singularity, since   
$F(\phi) + \frac{3}{2}\left( F'(\phi) \right)^2 = 1$  for such case. 
 
The Lagrangian (\ref{l2}) is ready to be considered under the Maupertuis-Jacobi
principle. For $V_{\rm eff}=0$, the Euler-Lagrange equations of (\ref{l2}) already correspond
to   geodesics of (\ref{m}). Moreover, from the energy constraint ${ \cal H}=0$, one has that
they in fact
correspond to timelike geodesics. For $V_{\rm eff}\ne 0$, let us introduce the
Jacobi (pseudo) metric
\beq
\label{jac}
\hat{G}_{AB} =  2|V_{\rm eff}| G_{AB}.
\eeq
One can see that, in accordance with 
  the Maupertuis-Jacobi principle, 
  the Euler-Lagrange equations of (\ref{l2}) correspond
to   geodesics of (\ref{jac}), parameterized by $s$, see Eq. (\ref{para}).  
From the energy constraint, one gets
\beq
\label{ener}
\hat G_{AB}\frac{d \phi}{ds}^A \frac{d\phi}{ds}^B =  -\frac{V_{\rm eff}}{ |V_{\rm eff}|},
\eeq
implying that the geodesic are timelike and spacelike for, respectively, 
$V_{\rm eff}>0$ and $V_{\rm eff}<0$. 
All the results\cite{townsend} that have motivated the present work can be obtained 
in a simpler way by
setting $F=1$. The analysis corresponding to the $\kappa=0$ case has already appeared in \cite{luciana}.

\section{The anisotropic case}

Our analysis can be extended in order to include also   anisotropic models. 
Let us consider a model
with action (\ref{act1}) and a Bianchi~I type
$D$-dimensional metric
\beq
ds^2 = -dt^2 + \sum_{i=1}^{D-1} a^2_i(t) dx^i,
\eeq
for which the scalar curvature is given by
\beq
R = 2\left(\sum_{i=1}^{D-1}\left(\dot H_i + H^2_i\right) +  \sum_{i=1, j>i}^{D-1} H_iH_j\right),
\eeq
where $H_i = \dot a_i/a_i$.
Again, by integrating the action (\ref{act1}) by parts, we obtain the following Lagrangian
\beq
\label{l22} 
 L   = 
\left( \prod_{i=1}^{D-1}a_i \right) \left(
-2F(\phi)\sum_{i=1,j>i}^{D-1}H_iH_j 
 - \vphantom{\sum_{i=1,j>i}^{D-1}} 2\left(\sum_{i=1}^{D-1} H_i \right)  \dot{\phi}^\alpha\partial_\alpha F(\phi)   
   + G_{\alpha\beta}(\phi)\dot{\phi}^\alpha \dot{\phi}^\beta - 2V(\phi) \right) 
\eeq
on the $(N+D-1)$-dimensional configuration space spanned by $(a_i,\phi^\alpha)$. 
As in the previous section, we will restrict ourselves here to 
$D=4$ case, the $D$-dimensional one follows straightforwardly and without
surprises. By introducing the following   metric
\beq
\label{m2}
G_{AB} = \left(
\begin{array}{cccc}
 0 & -a_3F & -a_2F & -a_2a_3\partial_\beta F  \\ 
 -a_3F & 0 &  -a_1F & -a_1a_3\partial_\beta  F  \\ 
 -a_2F &  -a_1F & 0 &  -a_1a_2\partial_\beta  F  \\ 
 -a_2a_3\partial_\alpha F & -a_1a_3\partial_\alpha F  & -a_1a_2\partial_\alpha F  & a_1a_2a_3 G_{\alpha\beta}
\end{array}  \right)
\eeq
on the configuration space (upper case roman indices run now over $1\dots N+3$), 
the Lagrangian (\ref{l22}) can be 
cast in the form given by (\ref{l2}), with $\phi^A= (a_1,a_2,a_3,\phi^\alpha)$ and
\beq
\label{aaa}
V_{\rm eff}(\phi^A) = a_1a_2a_3V(\phi^\alpha). 
\eeq
Note that the (pseudo) metric (\ref{m2}) has signature $(3,N)$. As in the isotropic case,
the determinant $\det G_{AB}$ plays a central role in the geometrical and dynamical analyses.
For $N=1$ and $G=1$, it reads
\beq
\det G_{AB} = -2(a_1a_2a_3F(\phi))^2\left(F(\phi) + \frac{3}{2}\left( F'(\phi) \right)^2 \right).
\eeq
 In this case, in addition to the
singularity present for the isotropic models, we have   also a singularity for 
$F(\phi)=0$. Such anisotropic singularity has been already described in \cite{sing}
in a dynamical analysis, where it is shown that it can be indeed used to rule out
large classes of models. We notice, nevertheless, that the dynamical analysis of
\cite{sing}  is considerably more involved than the geometrical analysis presented here.

As in the isotropic case, with the introduction of the metric (\ref{m2}) and the
effective potential (\ref{aaa}), the system is ready to be considered under the
Maupertuis-Jacobi principle.

\section{Discussion}

The great value of the dynamical analysis comes from the possibility of
ruling out large classes of models that are not viable from the theoretical point of view.
As we have already said,
 any candidate to a realistic cosmological model must exhibit certain qualitative 
 dynamical behaviors. For instance, the singularities corresponding to the condition
 $\det G_{AB}=0$ render the associated cosmological model inviable. Geometrically,
 such singularities imply that no geodesic can be extended beyond the singular points,
 what would imply a kind of future singularity in the associated cosmological model.
 This is precisely the case of the isotropic singularity of Section II and the
 anisotropic one of Section III. However, we stress that the identification of such
 singularities in the dynamical analyses\cite{sing1,sing} is considerably more involved than in the
 present geometrical approach. For multiple scalar fields, the dynamical analysis is
 much harder (see, for instance, \cite{double}), 
 and the present geometrical approach can be even more useful.
 
 The far most common dynamical analysis in cosmology is the stability classification
 of certain solutions. 
 De Sitter asymptotically stable fixed points are particularly relevant to
 the description of the recent accelerated expansion phase of the universe\cite{Peebles}.
 Such fixed points correspond to isotropic limits such that $\phi^\alpha$ and $H$ are
 constants, $\phi^A(s)=(a_0e^{Ht(s)},\phi_0^\alpha)$. 
 The geometrical analysis presented here can   help the identification of
 such  points. For this purpose, let us consider the the geodesic
 deviation equation, which governs the local tendency of nearby geodesics
 to converge or to diverge from each order
 \beq
 \label{gde}
 \dot\phi^A{\nabla}_A \dot\phi^B{\nabla}_B n^D = R_{ABC}^{\phantom{ABC}D}n^A\dot\phi^B\dot\phi^C,
 \eeq
 where $R_{ABC}^{\phantom{ABC}D}$ is the
     curvature tensor of the Jacobi metric,  and $n^A$ is a vector orthogonal to
   $\dot\phi^A$, pointing to 
   the direction of the deviation. Let us take, for instance, the same case of Section II,  $N=1$ and $G=1$, with
   $V_{\rm eff}(\phi)>0$ now. For the geodesic corresponding to a de Sitter fixed point, one has 
   $\dot\phi^A=((a_0H/ 2V_{\rm eff})e^{Ht},0)$. The only non vanishing component of (\ref{gde}) reads
\beq
\label{gde1}
\ddot{n}(s) = -R_{121}^{\phantom{121}2} {n}(s)\left(\dot\phi^{(1)}\right)^2  ,
\eeq
The stability of the de Sitter solution requires a bounded $n(s)$ for 
$s\rightarrow \infty$ and, hence,
$R_{121}^{\phantom{121}2}$ cannot be negative, eliminating, in this way, 
the possibility of finding out viable de Sitter fixed points inside regions where
$R_{121}^{\phantom{121}2}<0$. In the present case, one has
\beq
R_{121}^{\phantom{121}2}  =  \frac{3a F}{2V_{\rm eff}}
\left(  \frac{\Box V_{\rm eff}}{V_{\rm eff}}  -
\frac{\partial_a V_{\rm eff}\partial^aV_{\rm eff}}{V_{\rm eff}^2} 
\vphantom{\frac{FF'' - \frac{1}{2}\left(F'\right)^2}{a^3\left(F(\phi) + 
 \frac{3}{2}\left( F'(\phi) \right)^2 \right)^2}}  
 - 
\frac{FF'' - \frac{1}{2}\left(F'\right)^2}{a^3\left(F(\phi) + 
 \frac{3}{2}\left( F'(\phi) \right)^2 \right)^2}
\right).  
\eeq
For $F=1$, we have $R_{121}^{\phantom{121}2} = 3V''(\phi_0)/(2a^5V^2(\phi_0))$ on 
the fixed point $\phi_0$, and the usual requirement of stability $V''(\phi_0)>0$ is recovered.
For more general cases, the criteria $R_{121}^{\phantom{121}2}>0$ and $<0$ can be used to  
locate regions where  de Sitter fixed points could and could not appear, respectively.
We finish by noticing that our analysis is in perfect agreement with the comprehensive study
of \cite{fara}.

 \acknowledgments
 The authors thank Dr. Ricardo Mosna for valuable help, and FAPESP and CNPq for   financial
 support.

\end{document}